\documentclass{article}

\usepackage{comment}
\usepackage{hyperref}
\usepackage{amsmath}
\usepackage{amsfonts}
\usepackage{a4}
\usepackage[margin=1in]{geometry}
\usepackage{algpseudocode}

\def\eqref#1{(\ref{#1})}

\def\AddHere/{{\begin{center}\fbox{\Huge ADD SOMETHING HERE !!!}\end{center}}}

\newcommand{\RuleA}{\stackrel{A}{\Longrightarrow}}
\newcommand{\RuleB}{\stackrel{B}{\Longrightarrow}}

\newcommand{\Ntup}[1]{\left\langle{#1}\right\rangle}

\newcommand{\PAREN}[1]{\left({#1}\right)}
\newcommand{\BRACK}[1]{\left[{#1}\right]}
\newcommand{\BRACE}[1]{\left\{{#1}\right\}}

\newcommand{\ar}[1]{\linepenalty=10000{\small$\xrightarrow{-#1}$}\linepenalty=10}

\newcommand{\nCr}[2]{{{#1} \choose {#2}}}

\newcommand{\An}[3]{
  d_{#2}^r +
  \sum_{k = 1}^{r - 1} \BRACK{
    10^k \nCr{r}{k} D_{#1}^k d_{#2}^{r - k}
  } + 10^{-r} A_{#3}
}

\title{A Spigot-Algorithm for Square-Roots: \\
  Explained and Extended}
\author{Mayer Goldberg~(\href{mailto:gmayer@cs.bgu.ac.il}{gmayer@cs.bgu.ac.il})\footnote{Department of Computer Science, Ben-Gurion University of the Negev, PO Box 653, Beer Sheva 8410501, Israel.}}

\begin{document}\maketitle\raggedright

\begin{abstract}
  \noindent This work presents and extends a known spigot-algorithm for computing square-roots, digit-by-digit, that is suitable for calculation by hand or an abacus, using only addition and subtraction. We offer an elementary proof of correctness for the original algorithm, then present a corresponding spigot-algorithm for computing cube-roots. Finally, we generalize the algorithm, so as to find $r$-th roots, and show how to optimize the algorithm for any $r$. The resulting algorithms require only integer addition and subtraction. 
\end{abstract}

\section{Introduction}
This paper explores a curious spigot-algorithm for computing square roots, and then extends it to computing $r$-th roots. The original algorithm, believed to have originated in Japan, is particularly suitable to be carried out either by hand or using an abacus.

Spigot-algorithms are different from other approximation algorithms: Whereas the output of approximation algorithms is a better approximation, the output of a spigot-algorithm is the subsequent digit(s) of the result, in some decimal or other notation, as well as any additional data needed for subsequent iterations of the algorithm. 

There is something particularly satisfying about spigot-algorithms, such as the algorithms for computing quotients by long division, or for computing logarithms~\cite{Goldberg2006}: A spigot-algorithm is tightly coupled to a specific, nested, or layered representation of the result, and each iteration of the algorithm ``peels off a layer'' from this representation, extracting a digit in turn. 

The search for such spigot-algorithms for extracting square-roots led me to the website of Prof~Frazer Jarvis, of the mathematics department at the University of Sheffield. Prof Jarvis presented an algorithm that was introduced to him in school as ``a Japanese algorithm'', and which I found to be marvelously simple:

To find the root of a number $M$, start with the pair $\Ntup{5 M, 5}$. The algorithm constructs a sequence of pairs of integers, according to two rules, to which we shall refer as Rule~A and Rule~B:

\begin{description}
\item [Rule~A: ] For the pair $\Ntup{P, Q}$, if $P \geq Q$, the next element in the sequence is the pair $\Ntup{P - Q, Q + 10}$.

\item [Rule~B: ] For the pair $\Ntup{P, Q}$, if $P < Q$, the next element in the sequence is the pair $\Ntup{100P, 10Q - 45}$. 
\end{description}

Note that because $Q$ is a multiple of 5, $10Q - 45$ can be computed by simply
inserting a 0 between the right-most digit of $Q$ and the digit to its left. 

The number of applications of Rule~A between successive applications
of Rule~B is the subsequent digit in the square root of $M$. Here's an
example: 

We wish to compute the square root of $3$. Starting with the pair
$\Ntup{15, 5}$, we proceed as follows:

\begin{quote}
  $\Ntup{15, 5} \RuleA \Ntup{10, 15} \RuleB \Ntup{1000, 105} \RuleA
  \Ntup{895, 115} \RuleA \Ntup{780, 125} \RuleA \Ntup{655, 135} \RuleA
  \Ntup{520, 145} \RuleA \Ntup{375, 155} \RuleA \Ntup{220, 165} \RuleA
  \Ntup{55, 175} \RuleB \Ntup{5500, 1705} \RuleA \Ntup{3795, 1715}
  \RuleA \Ntup{2080, 1725} \RuleA \Ntup{355, 1735} \RuleB \Ntup{35500,
    17305} \RuleA \Ntup{18195, 17315} \RuleA \Ntup{880, 17325} \RuleB
  \Ntup{88000, 173205} \RuleB \Ntup{8800000, 1732005} \RuleA
  \Ntup{7067995, 1732015} \RuleA \Ntup{5335980, 1732025} \RuleA
  \Ntup{3603955, 1732035} \RuleA \Ntup{1871920, 1732045} \RuleA
  \Ntup{139875, 1732055} \RuleB \cdots $

\end{quote}
The number of applications of Rule~A between successive applications
of Rule~B are $1, 7, 3, 2, 0, 5, \ldots$, and hence the square root of
$3$ is given by $1.73205\cdots$. Each digit is correct; no round-off errors are involved. If we can continue this algorithm with pairs of arbitrarily-large
integers, we can compute the square root to arbitrarily-many digits.

Neither this algorithm, nor any other spigot algorithm for extracting square-roots, are a match for the efficiency of $O(n^2)$ algorithms, such as
Newton-Raphson, etc. However, in comparison with other spigot-algorithms, this one stands out in several ways:
\begin{itemize}
\item This algorithm uses only the most elementary operations:
  Addition, subtraction, multiplication by a power of the counting
  base (10), and a single multiplication by 5, during what appears to
  be an initialization of a computation. In contrast with this
  simplicity, other spigot-algorithms involve many more operations,
  including the computationally-expensive, unrestricted use of integer
  multiplication. 

\item This algorithm arrives at its answer by the unusual route of
  counting the number of times a specific rule is applied, between
  subsequent applications of another rule. This number of applications
  of a rule is not used to control the algorithm, but rather this
  number \textit{is} the next digit in the answer. 
\end{itemize}
Contacting Prof~Jarvis~\cite{jarvis-personal}, I learned that he was taught this algorithm in school, and that his teacher claimed it was a Japanese algorithm. While I was not able to ascertain either the name of its creator, or the time frame in which it was created, it seems plausible that this algorithm was created by an abacus user: Addition, subtraction, and multiplication by a power of 10 (shifting), are all elementary operations on an abacus.

Prof~Jarvis did post on his website a proof of correctness for this algorithm. The proof, however, relies on calculus, and shows that the resulting number converges to the desired square-root. While the proof is certainly correct, it makes free use of modern calculus, and does not appear to be an argument that would have been made by a Japanese mathematician working several hundreds of years ago. Nor does it uncover the nested, layered representation that I expected to find in a spigot-algorithm. So this is the starting-point of my inquiry into this ``Japanese algorithm''. My goals were:
\begin{itemize}
\item To uncover the layered representation that is at the heart of this algorithm, in order to understand what insights it captured about square-roots.

\item To use this layered representation to understand how the algorithm works.

\item To generalize the algorithm to other roots.
\end{itemize}

The rest of this paper reports my findings.

\section{Notation}

We introduce the notation
\begin{eqnarray*}
  [d_0; d_1; \cdots; d_n]
  & = & 10^n d_0 + 10^{n - 1} d_1
  + 10^{n - 2} d_2 + \cdots + d_n
\end{eqnarray*}
to denote the natural number, in decimal notation, consisting of the digits $d_0, \ldots, d_n$, with $d_0$ being the coefficient of the highest power of 10. When this won't lead to ambiguities, we may abbreviate $[d_0; d_1; \cdots; d_n]$ further, and just write $D_n$. Note that
\begin{eqnarray*}
  [d_0; d_1; \cdots; d_n; d_{n + 1}]
  & = & 10 [d_0; d_1; \cdots; d_n] + d_{n + 1} \\
  & = & 10 D_n + d_{n + 1}
\end{eqnarray*}

\section{Square Roots}
\label{sec:sqrt}
The square-root of the non-negative real number $M$ is a real number given by the infinite decimal expansion
\begin{eqnarray*}
  \sqrt M & = & \sum_{k = 0}^{\infty} 10^{-k} d_k
\end{eqnarray*}
We require that $0 \leq d_k \leq 9$ for all $k \geq 0$. 

Our problem is, given $M$, to extract the digits $d_k$, spigot-like, that is, one digit at a time, for $k = 0, 1, 2, \ldots$.

First, we explore a suitable representation for the number $M$, that enables a digit-by-digit extraction of its root:

\begin{eqnarray*}
  M & = &
  \PAREN{
    d_0 +
    \frac{d_1}{10} +
    \frac{d_2}{10^2} +
    \frac{d_3}{10^3} +
    \cdots}^2 \\
  & = & d_0^2 + \frac1{100} A_1 \mbox{, where $A_1$ is given by} \\
  A_1 & = & d_1^2 + 20 d_1 D_0
  + \frac1{100} A_2 \mbox{, where $A_2$ is given by} \\
  A_2 & = & d_1^2 + 20 d_2 D_1 
  + \frac1{100} A_3 \mbox{, etc} \\
  & \cdots & \\
  A_{n + 1} & = & \mbox{\fbox{$d_{n + 1}^2 + 20 d_{n + 1} D_n$}}
  + \frac1{100} A_{n + 2} \mbox{, etc} 
\end{eqnarray*}

Note that these equations specify an arbitrarily-nested expression, where each $A_{n + 1}$ is nested within $A_n$.

Our problem becomes to compute the \fbox{boxed sub-expression} that is part of $A_{n + 1}$, and specifically, to find $d_{n + 1}$. Then we can subtract it from $A_{n + 1}$, multiply by 100, and repeat with $A_{n + 2}$ and $d_{n + 2}$. Consider the following sum:

\begin{eqnarray}
  \lefteqn{\left\{\begin{array}{l}
    2 [d_0; \cdots ; d_n; 0]^2 + 1 \nonumber\\[1.0em]
    +\ 2 [d_0; \cdots ; d_n; 1]^2 + 1 \nonumber\\[1.0em]
    +\ 2 [d_0; \cdots ; d_n; 2]^2 + 1 \nonumber\\[1.0em]
    \cdots \nonumber\\[1.0em]
    +\ 2 [d_0; \cdots ; d_n; d_{n+1} - 1]^2 + 1 \nonumber\\[1.0em]
    \end{array}\right.} \\
  & = & \sum_{k = 0}^{d_{n + 1} - 1} \BRACK{
    2 (10 D_n + k)^2 + 1
  } \label{eq:sum:sqrt} \\
  & = & \mbox{\fbox{$d_{n + 1}^2 + 20 d_{n + 1} D_n$}} \nonumber
\end{eqnarray}

Notice that $[d_0; d_1; \cdots; d_n]$ is the number represented by the initial $n$ digits within the $n+1$-digit number $[d_0; d_1; \cdots; d_n; d_{n + 1}]$. Also notice that the value of this sum is precisely the boxed sub-expression in $A_{n + 1}$: 

\begin{eqnarray*}
  A_{n+1} & = & \mbox{\fbox{$d_{n+1}^2 + 20 d_{n+1} D_n$}}
  + \frac1{100} A_{n+2} \\
\end{eqnarray*}  

We skipped a few steps in evaluating the summation, but this essentially amounts to an application of the well known formulae for sums of powers:

\begin{eqnarray*}
  \begin{array}{rclrl}
    S_0(n) & = & \sum_{k = 0}^n 1 & =& n \\[1em]
    S_1(n) & = & \sum_{k = 0}^n k & =& \frac12 n^2 + \frac12 n \\
  \end{array}
\end{eqnarray*}

So the idea is to subtract from $A_{n + 1}$ subsequent terms of the form $[d_0; \cdots; d_n; k]$, for $k = 0, 1, ...$, from the Equation~\eqref{eq:sum:sqrt}, leaving what we call a \textit{remainder}.

The upper-limit of the sum is $d_{n + 1} - 1$, which makes it appear as if need to know the value of $d_{n + 1}$ in order to know how many summands to take, namely, $d_{n + 1}$, but this is not so: The \textit{next} remainder $\frac1{100} A_{n + 2}$ is sufficiently small, relative to the current remainder, that we can keep subtracting $[d_0; \cdots; d_n; k]$, for $k = 0, 1, ... d_{n + 1} - 1$, from the \textit{current} remainder, \textit{until the difference is negative}. 

\vskip1em
\noindent\textbf{Example: } We wish to extract the first four digits of $\sqrt 7$:
\begin{enumerate}
\item $7$ \ar{(2 \cdot 0 + 1)} $6$ \ar{(2 \cdot 1 + 1)} $\mathbf{3}$ \ar{(2 \cdot 2 + 1)} $-2 < 0$
\item $\mathbf{3}00$ \ar{(2 \cdot 20 + 1)} $259$ \ar{(2 \cdot 21 + 1)} $216$ \ar{(2 \cdot 22 + 1)} $171$ \ar{(2 \cdot 23 + 1)} $124$ \ar{(2 \cdot 24 + 1)} $75$ \ar{(2 \cdot 25 + 1)} $\mathbf{24}$ \ar{(2 \cdot 26 + 1)} $-29 < 0$
\item $\mathbf{24}00$ \ar{(2 \cdot 260 + 1)} $1879$ \ar{(2 \cdot 261 + 1)} $1356$ \ar{(2 \cdot 262 + 1)} $831$ \ar{(2 \cdot 263 + 1)} $\mathbf{304}$ \ar{(2 \cdot 264 + 1)} $-225 < 0$
\item $\mathbf{304}00$ \ar{(2 \cdot 2640 + 1)} $25119$ \ar{(2 \cdot 2641 + 1)} $19836$ \ar{(2 \cdot 2642 + 1)} $14551$ \ar{(2 \cdot 2643 + 1)} $9264$ \ar{(2 \cdot 2644 + 1)} $\mathbf{3975}$ \ar{(2 \cdot 2645 + 1)} $-1361 < 0$
\item $\mathbf{3975}00$ \ar{(2 \cdot 26450 + 1)} $\cdots$
\end{enumerate}

If $M$ is given by several digits, we divide $M$ into groups of \textit{two digits} each, starting \textit{from the right}. Multiplying by 100 is the same as appending the next, \textit{leftmost} group of digits. So for example, the process extracting a square-root is the same for $\sqrt {2.3456}, \sqrt{234.56}, \sqrt{23456}, \sqrt{2345600}$, etc.

\subsection{Optimizing the Algorithm For Square-Roots}
\label{ssec:simp}

In the previous section, we subtracted terms of the form
\begin{eqnarray}\label{eq:sqseq}
  2 [d_0; \cdots; d_n; j] + 1 \mbox{, for $j = 0, 1, \ldots, d_{n + 1}$}
\end{eqnarray}
from $A_{n + 1}$. We can compute $2 [d_0; \cdots; d_n; j] + 1$ for successive values of $j$ by considering the difference-sequence, which is the constant 2. So starting with $2 [d_0; \cdots; d_n; 0] + 1$, we can compute successive terms by adding 2 to the previous terms.

\subsection{Reconstructing the Original ``Japanese Algorithm''}
\label{ssec:sqrtdiff}

The ``Japanese algorithm'' is essentially the optimized algorithm of the previous section, with one additional, non-mathematical optimization, that is probably intended to simplify computation on an abacus.

Rather than start with $\Ntup{M, 1}$, and adding 2 to the remainder in each application of Rule~A, we multiply the initial setup by 5, by starting with $\Ntup{P, Q} = \Ntup{5M, 5}$, and add 10 in each application of Rule~A. Adding 10 on an abacus is a bit simpler than adding 2, because it just adds 1 a column to the tens-column on the abacus. An additional simplification is gained in Rule~B: Rather than changing the second tuple from $2 [d_0; \cdots; d_n] + 1$ to $2 [d_0; \cdots; d_n; 0] + 1$, which requires us to update $Q$ in the tuple $\Ntup{P, Q}$ with $10Q - 9$, we now update $Q$ with $10Q - 45$. \textit{Mathematically}, this isn't any simpler than $10Q - 9$, but on an abacus, which operates on the individual, decimal digits of numbers, this operation amounts to inserting a zero-digit just before the rightmost digit of $Q$, so it is very simple to update the value of $Q$ in place, on the abacus.

So if we are computing square-roots by hand, digit-by-digit, for fun, on the back of an envelope, this non-mathematical optimization is of little use to us, and merely obfuscates the algorithm in Section~\ref{ssec:sqrtdiff}. But if we are computing square-roots on an abacus, then the ``Japanese algorithm'' does simplify our manual task.

\section{Cube Roots}
\label{sec:croot}
We now turn to the problem of finding a corresponding spigot-algorithm for finding cube-roots. Suppose we wanted to compute the cube-root of $M$, given by the infinite decimal expansion
\begin{eqnarray*}
  \sqrt[3]{M} & = & \sum_{k = 0}^{\infty} 10^{-k} d_k
\end{eqnarray*}
The nested expression for $M$ would be
\begin{eqnarray*}
  M
  & = & \PAREN{
    d_0 +
    \frac{d_1}{10} +
    \frac{d_2}{10^2} +
    \frac{d_3}{10^3} +
    \cdots }^3
  \\
  & = & d_0^3 + \frac1{1000} A_1 \mbox{, where $A_1$ is given by} \\
  A_1 & = & d_1^3 + 30 d_1 D_0 D_1 +
  + \frac1{1000} A_2 \mbox{, where $A_2$ is given by} \\
  A_2 & = & d_2^3 + 30 d_2 D_1 D_2 +
  + \frac1{1000} A_3 \mbox{, etc} \\
  & \cdots & \\
  A_{n + 1} & = & d_{n + 1}^3 + 30 d_{n + 1} D_n D_{n + 1}
  + \frac1{1000} A_{n + 2} \mbox{, etc} \\
  & = & d_{n + 1}^3 + 30 d_{n + 1} D_n (10 D_n + d_{n + 1}) + \frac1{1000} A_{n + 2} \mbox{, etc} \\
  & = & \mbox{\fbox{$d_{n + 1}^3 + 30 d_{n + 1}^2 D_n + 300 D_n^2$}} + \frac1{1000} A_{n + 2} \mbox{, etc} 
\end{eqnarray*}

Just as before, this representation is nested, with $A_{n + 1}$ being a part of $A_n$, for all $n \in \mathbb{N}$. Our goal is to find the \fbox{boxed sub-expression}, and the respective digit $d_{n + 1}$, and subtract it from the current $A_{n + 1}$, multiply the remainder by $10^3$, and continue with $A_{n + 2}$. We consider the following sum:
\begin{eqnarray}
  \lefteqn{\left\{\begin{array}{l}
    3 [d_0; \cdots; d_n; 0]^2 + 3 [d_0; \cdots; d_n; 0] + 1 \nonumber\\[1.0em]
    +\ 3 [d_0; \cdots; d_n; 1]^2 + 3 [d_0; \cdots; d_n; 1] + 1 \nonumber\\[1.0em]
    +\ 3 [d_0; \cdots; d_n; 2]^2 + 3 [d_0; \cdots; d_n; 2] + 1 \nonumber\\[1.0em]
    \cdots \nonumber\\[1.0em]
    +\ 3 [d_0; \cdots; d_n; d_{n + 1} - 1]^2
    +\ 3 [d_0; \cdots; d_n; d_{n + 1} - 1] + 1 \nonumber\\[1.0em]
    \end{array}\right.} \nonumber\\
  & = & \sum_{k = 0}^{d_{n + 1} - 1} \BRACK{
    3 (10 D_n + k)^2) +
    3 (10 D_n + k)
  } \label{eq:sum:cube}\\
  & = & d_{n + 1}^3 + 3 d_{n + 1} (10 D_n)
  + 3 (10 D_n)^2 \nonumber\\
  & = & \mbox{\fbox{$d_{n + 1}^3 + 30 d_{n + 1}^2 D_n + 300 D_n^2$}}
  \nonumber
\end{eqnarray}
The value of the sum is precisely the value of the boxed sub-expression in $A_{n + 1}$:
\begin{eqnarray*}
  A_{n + 1} & = & 
  \mbox{\fbox{$d_{n + 1}^3 + 30 d_{n + 1}^2 D_n + 300 D_n^2$}}
  + \frac1{1000} A_{n + 2} \mbox{, etc} 
\end{eqnarray*}
To complete this last sum, we used the formulae for $S_0(n), S_1(n)$, given earlier, as well as the formula for $S_2(n)$, given by:
\begin{eqnarray*}
  S_2(n) & = &
  \sum_{k = 0}^{n} k^2
  ~=~ \frac13 n^3 + \frac12 n^2 + \frac16 n
\end{eqnarray*}
Just as before, the upper-limit of the sum, $d_{n + 1} - 1$ is computed by repeatedly subtracting $[d_0; \ldots; d_n; k]$, for $k = 0, 1, \ldots$, from the current remainder, until the difference is negative.

\vskip1em
\noindent\textbf{Example: } We wish to extract the first four digits of $\sqrt[3]{7}$:
\begin{enumerate}
\item $7$ \ar{(3 \cdot 0^2 + 3 \cdot 0 + 1)} $\mathbf{6}$ \ar{(3 \cdot 1^2 + 3 \cdot 1 + 1)} -1
\item $\mathbf{6}000$ \ar{(3 \cdot 10^2 + 3 \cdot 10 + 1)} $5669$ \ar{(3 \cdot 11^2 + 3 \cdot 11 + 1)} $5272$ \ar{(3 \cdot 12^2 + 3 \cdot 12 + 1)} $4803$ \ar{(3 \cdot 13^2 + 3 \cdot 13 + 1)} $4256$ \ar{(3 \cdot 14^2 + 3 \cdot 14 + 1)} $3625$ \ar{(3 \cdot 15^2 + 3 \cdot 15 + 1)} $2904$ \ar{(3 \cdot 16^2 + 3 \cdot 16 + 1)} $2087$ \ar{(3 \cdot 17^2 + 3 \cdot 17 + 1)} $1168$ \ar{(3 \cdot 18^2 + 3 \cdot 18 + 1)} \ar{(3 \cdot 18^2 + 3 \cdot 18 + 1)} $\mathbf{141}$ \ar{(3 \cdot 19^2 + 3 \cdot 19 + 1)} $-1000$
\item $\mathbf{141}000$ \ar{(3 \cdot 190^2 + 3 \cdot 190 + 1)} $\mathbf{32129}$ \ar{(3 \cdot 191^2 + 3 \cdot 191 + 1)} $-77888$
\item $\mathbf{32129}000$ \ar{(3 \cdot 1910^2 + 3 \cdot 1910 + 1)} $21178969$ \ar{(3 \cdot 1911^2 + 3 \cdot 1911 + 1)} $\mathbf{10217472}$ \ar{(3 \cdot 1912^2 + 3 \cdot 1912 + 1)} $-755497$
\item $\mathbf{10217472}000$ \ar{(3 \cdot 19120^2 + 3 \cdot 19120 + 1)} $\cdots$
\end{enumerate}

If $M$ is given by several digits, we divide $M$ into groups of \textit{three} digits each, starting \textit{from the right}. Multiplying by $10^3 = 1000$ is the same as appending the next \textit{leftmost} group of digits. So for example, the process of extracting the cube-root is the same for $\sqrt[3]{2.3456}$, $\sqrt[3]{2345.6}$, $\sqrt[3]{2345600}$, etc.

\subsection{Optimizing the Algorithm for Cube-Roots}

In Equation~\eqref{eq:sum:cube}, we subtracted terms of the form
\begin{eqnarray*}
3 [d_0; \cdots; d_n; j]^2 + 3 [d_0; \cdots; d_n; j] + 1
\end{eqnarray*}
from $A_{n + 1}$. In Section~\ref{ssec:simp}, we make use of difference-sequences to compute sum-up the terms of the sequence~\eqref{eq:sqseq} efficiently. Because terms in the sequence are quadratic, we require two levels of difference-sequences, that is a difference-sequence, and the difference-sequence of the difference-sequence, in order to eliminate all non-trivial multiplications, that is, all multiplications not by a power of 10 (the counting base).

After some algebraic manipulation, our algorithm for cube-roots becomes:

\noindent\textbf{Initial setup:} To find $\sqrt[3]{M}$, we set up the quintuple $\Ntup{M, D = 0, R = 0, S = 0, W = 1}$.

\noindent\textbf{Rule~A:} When $M \geq W$:
\begin{eqnarray*}
  \Ntup{M, D, R, S, W} & \RuleA &
  \Ntup{M - W, D + 1, R + 1, S', W + S'} \mbox{, where} \\
  S' & = & S + 6
\end{eqnarray*}

\noindent\textbf{Rule~B:} When $M < W$, the next digit is $D$, and:
\begin{eqnarray*}
  \Ntup{M, D, R, S, W} & \RuleB &
  \Ntup{1000M, 0, 10R, 10S, 100W - 270R - 99}
\end{eqnarray*}

We have not managed to avoid multiplying by the constant 270 in the expression $100W - 270R - 99$, but fortuitously, we can compute this expression relatively painlessly on an abacus by re-arranging it as: $100W - 300R + 30R - 99$, thrice subtracting $R$ from the third column from the right, and thrice adding $R$ to the second column from the right. Similarly, subtracting $99$ can be done quickly by adding $100$ and subtracting $1$, which is just adding $1$ to the third column from the right, and subtracting $1$ from the rightmost column.

\section{The General Meta-Algorithm}

The nested, onion-like representation for computing the $r$-th root requires that we shift by $n$ digits at a time, that is, multiply by $10^r$. This quickly becomes unwieldy for manual computation, but it can certainly be used to extract digits using a computer program. We call this a meta-algorithm, because just as with the algorithms for computing square-roots and cube-roots, it will involve many non-trivial multiplications. To make this a practical algorithm for computing by hand or using an abacus would require that we eliminate all powers from the computation, so we would need to use $r - 1$ repeated difference-sequences and then worry about how to optimize the manual computation. So we settle for calling this a meta-algorithm, suggesting that for any value of $r$, this is an algorithm for generating algorithms for manually computing the $r$-th root, digit-by-digit.

Suppose we wanted to compute the $r$-th root of $M$, given by the infinite decimal expansion
\begin{eqnarray*}
  \sqrt[r]{M} & = & \sum_{k = 0}^{\infty} 10^{-k} d_k
\end{eqnarray*}

The nested expression for $M$ would be

\begin{eqnarray*}
  M & = & 
  \PAREN{
    d_0 +
    \frac{d_1}{10} +
    \frac{d_2}{10^2} +
    \frac{d_3}{10^3} +
    \cdots }^r \\
  & = & d_0^r + 10^{-r} A_1 \mbox{, where} \\
  A_{1} & = & \An{0}{1}{2} \\
  A_{2} & = & \An{1}{2}{3} \\
  & \cdots & \\
  A_{n + 1} & = & \An{n}{n+1}{n+2} \\
  & = & \mbox{\fbox{$(10 D_n + d_{n + 1})^r - (10 D_n)^r$}}
  + 10^{-r} A_{n + 2}
\end{eqnarray*}
Just as before, this representation is nested, with $A_{n + 1}$ being a part of $A_n$, for all $n \in \mathbb{N}$. Our goal is to find the \fbox{boxed sub-expression}, and the respective digit $d_{n + 1}$, and subtract it from the current $A_{n + 1}$, multiply the remainder by $10^r$, and continue with $A_{n + 2}$. We consider the following sum:
\begin{eqnarray*}
  \lefteqn{\sum_{j = 0}^{10 D_n + d_{n + 1} - 1} \BRACK{
      \sum_{k = 1}^{r} \nCr{r}{k} j^{r - k}
  }} \\
  & = &
  \sum_{j = 1}^{r} \BRACE{
    \nCr{r}{j} \BRACK{
      \PAREN{
        \sum_{k = 0}^{10 D_n + d_{n + 1} - 1} k^{r - j}
      }
      -
      \PAREN{
        \sum_{k = 0}^{10 D_n - 1} k^{k - j}
      }
    }
  } \\
  & = &
  \BRACE{
    \sum_{j = 0}^{10D_n + d_{n + 1} - 1}
    \PAREN{\sum_{k = 1}^{r} \BRACK{\nCr{r}{k} j^{r - k}}}
  }
  -
  \BRACE{
    \sum_{j = 0}^{10D_n - 1}
    \PAREN{\sum_{k = 1}^{r} \BRACK{\nCr{r}{k} j^{r - k}}}
  } \\
  & = &
  \BRACE{
    \sum_{k = 0}^{10 D_n + d_{n + 1} - 1} \BRACK{
      (k + 1)^r - k^r
    }
  } -
  \BRACE{
    \sum_{k = 0}^{10 D_n - 1} \BRACK{
      (k + 1)^r - k^r
    }
  } \\
  & = &
  \mbox{\fbox{$(10 D_n + d_{n + 1})^r - (10 D_n)^r$}}
\end{eqnarray*}
The value of the sum is precisely the value of the boxed sub-expression in $A_{n + 1}$:
\begin{eqnarray*}
  A_{n + 1}
  & = & \mbox{\fbox{$(10 D_n + d_{n + 1})^r - (10 D_n)^r$}}
  + 10^{-r} A_{n + 2}
\end{eqnarray*}

Just as we have done in Section~\ref{sec:sqrt} and Section~\ref{sec:croot}, we find $d_{n + 1}$ by repeatedly subtracting terms of the form
\begin{eqnarray*}
\sum_{k = 1}^{r} \BRACK{\nCr{r}{k} [d_0; \cdots; d_n; j]^k}
\end{eqnarray*}
for $j = 0, 1, \ldots$, from the current remainder, until the difference is negative. At that point, we have found the upper limit $d_{n + 1} - 1$ of the sum
\begin{eqnarray*}
  \sum_{j = 0}^{d_{n + 1} - 1} \BRACE{
    \sum_{k = 1}^{r} \BRACK{
      \nCr{r}{k} [d_0; \cdots; d_n; j]^k}}
\end{eqnarray*}
have computed it from $A_{n + 1}$, and have computed the remainder, with which we can continue the computation.

All these summations may seem overly complicated, but the meta-algorithm itself is very simple, so a few steps of a worked-out example should help clear matters: We would like to find the first few digits of $\sqrt[5]{7}$. Because we are computing the fifth power, we shall need to compute the function $f(n) = (n + 1)^5 - n^5 = 1 + 5 n + 10 n^2 + 10 n^3 + 5 n^4$ at each step, for various values of $n$. Because this is the meta-algorithm, we are not going to optimize this step using difference series, but of course, this is straightforward, even if tedious.

\begin{enumerate}
\item $7$ \ar{f(0)} $\mathbf{6}$ \ar{f(1)} $-25 < 0$
\item $\mathbf{6}00000$ \ar{f(10)} $538949$ \ar{f(11)}
  $451168$ \ar{f(12)} $328707$ \ar{f(13)}
  $\mathbf{162176}$ \ar{f(14)} $-59375 < 0$
\item $\mathbf{162176}00000$ \ar{f(140)}
  $14269163299$ \ar{f(141)}
  $12264660768$ \ar{f(142)}
  $10202891057$ \ar{f(143)}
  $8082635776$ \ar{f(144)}
  $5902659375$ \ar{f(145)}
  $3661709024$ \ar{f(146)}
  $\mathbf{1358514493}$ \ar{f(147)}
  $-1008211968 < 0$
\item $\mathbf{1358514493}00000$ \ar{f(1470)}
  $112472218403649$ \ar{f(1471)}
  $89029327290368$ \ar{f(1472)}
  $65522646041407$ \ar{f(1473)}
  $41952044561376$ \ar{f(1474)}
  $\mathbf{18317392578125}$ \ar{f(1475)}
  $-5381440357376 < 0$
\end{enumerate}
So far, we managed to extract $1.475$ so far.

To find the $r$-th root, we simply define $f(n) = (n + 1)^r - n^r$, expand, simplify, and then use difference series to make the computation of successive values of $f(n)$ more efficient.

\subsection{Programming the Meta-Algorithm}

As noted previously, the meta-algorithm cannot be optimized for all $r$. Rather, for each $r$, a different optimization exists by means of $r - 1$ nested, difference-sequences.

It is straightforward, however, to program the meta-algorithm, as-is, in its un-optimized form, in any of the ever-growing list of programming-language that offer support for arbitrary-precision integer-arithmetic (e.g., Javascript, Python, LISP, dc, and many others).

We implemented the meta-algorithm in Scheme, which is a quasi-functional programming-language of the LISP-variety. The meta-algorithm is implemented as the procedure \verb|root|, which takes 3 arguments: The integer $r$, the integer $M$, and the number of digits $d$, and returns the first $d$ digits of the $\sqrt[r]{M}$, the $r$-th root of $M$. No decimal point is specified, and the algorithm returns the same result for the $r$-th root of $M \cdot 10^{rk}$, for any natural number $k$. Some of our implementations returns the result as a list of $d$ digits, while other implementations return a single $d$-digit number. This is more a matter of convenience and whether the underlying programming language has native support for arbitrarily-long integers (called \textit{bignums} in LISP, \textit{bigint} in JavaScript, \textit{LargeInteger} in Java, etc). Below is an example of computing $\sqrt[5]{7}$ to $201$ decimal places:
\begin{verbatim}
> (root 5 7 201)
14757731615945520692769166956322441065440
 9361374020356777090416888452176749920836
 0714411082351298307654442294189726695499
 1677818301896039335532935966839393186145
 4579258848931485233873464556602592552045
\end{verbatim}
So $\sqrt[5]{7}$ starts with the digits:
\begin{eqnarray*}
  1.\!\!\!\begin{array}[t]{l@{}}
  47577316159455206927691669563224410654409361374020 \\
  35677709041688845217674992083607144110823512983076 \\
  54442294189726695499167781830189603933553293596683 \\
  93931861454579258848931485233873464556602592552045\cdots \\
  \end{array}
\end{eqnarray*}
No rounding is performed, and the digits are extracted one at a time, so if $n$-digits are desired, it is best to generate $n + 1$ digits and then decide how to round-off the $n$-th digit.

\subsection{Non-Ineteger Roots}

The nested, layered representation we found for the $r$-th root is:
\begin{eqnarray*}
  A_{n + 1}
  & = & d_{n + 1}^r +
  \sum_{j = 0}^{10D_n + d_{n + 1} - 1} \left[
    \sum_{k = 1}^{r} \nCr{r}{k}j^{r - k}
    \right] +
  10^{-r}A_{n + 2} \\
  & = & \mbox{\fbox{$(10D_n + d_{n + 1})^r - (10D_n)^r$}} + 10^{-r}A_{n + 2}
\end{eqnarray*}
We have seen how the subsequent digit $d_{n + 1}$ is found using only $d_0, \ldots, d_n$. This holds both for the finite and infinite (aka \textit{extended}) version of the binomial theorem:
\begin{eqnarray*}
  (x + y)^{\alpha} & = &
  \sum_{k \leq \alpha}^{\alpha} \nCr{\alpha}{k}x^ky^{\alpha - k}
\end{eqnarray*}
for $\alpha \in \mathbb{R}$.

Our meta-algorithm is impractical as a spigot-algorithm for finding non-integer roots, because using it requires us to raise an integer to a non-integer power, and to do so many times, is not practical as a spigot algorithm. However, the underlying mathematical relationship holds, of course.

For example, we wish to compute the first few digits of $\sqrt[3.14]{2.71}$. We define the function $f(x) = (x + 1)^{3.14} - x^{3.14}$. We then proceed as usual, \textit{mutatis mutandis}, namely, upon computing another digit, we shall have to multiply by a non-integer power of ten, 

\begin{enumerate}
\item $2.71$ \ar{f(0)} $\mathbf{1}$ \ar{f(1)} $-6.10524092701 < 0$
\item $\mathbf{1} \cdot 10^{3.14}$ \ar{f(10)} $1878.86984778$ \ar{(f11)}
  $1293.86872199$ \ar{f(12)} $\mathbf{594.671786845}$ \ar{f(13)}
  $-229.629676871 < 0$
\item $\mathbf{594.671786845} \cdot 10^{3.14}$ \ar{f(130)} $715111.365835$
  \ar{f(131)} $607605.215136$ \ar{f(132)} $498341.958142$ \ar{f(133)}
  $387306.411836$ \ar{f(134)} $274483.377207$ \ar{f(135)} $159857.639358$
  \ar{f(136)} $\mathbf{43413.9676045}$ \ar{f(137)} $-74862.8844241 < 0$
\item $\mathbf{43413.9676045} \cdot 10^{3.14}$ \ar{f(1370)} $43715426.8423$
  \ar{f(1371)} $27477569.9671$ \ar{f(1372)} $\mathbf{11214366.0602}$
  \ar{f(1373)} $-5074205.94065 < 0$
\end{enumerate}
And so we find that the first digits of $\sqrt[3.14]{2.71} = 1.373\cdots$, and we can continue this computation for a bit more. But the one major difference about non-ineteger roots is that the computation of $(x + 1)^{\alpha} - x^{\alpha}$ cannot be carried out practically with arbitrary precision.

\section{Discussion \& Conclusion}

In this work, we presented a curious ``Japanese algorithm'' for computing square roots, and generalized it to cube roots, and beyond.

The origins of this algorithm remain a mystery to me and a subject of conjecture. The algorithm seems to have been adjusted for easy computation on an abacus. The abacus was introduced to Japan in the 14th century, and came to be used widely in the 17th century. It was since the 17th century that Japanese mathematicians studied the abacus and either adapted or created algorithms for its use. The most well-known of these Japanese mathematicians was \textit{Seki Takakazu}~(1642--1708). Seki worked in many areas of mathematics, including algebra. He knew the binomial theorem, Horner's method for finding square roots, and had independently discovered Bernoulli numbers. He was also interested in finding roots of various functions. It is quite reasonable that this ``Japanese algorithm'' was discovered either by him or by one of his pupils. 

From the equation
\begin{eqnarray*}
  A_{n + 1}
  & = & d_{n + 1}^r +
  \sum_{j = 0}^{10D_n + d_{n + 1} - 1} \left[
    \sum_{k = 1}^{r} \nCr{r}{k}j^{r - k}
    \right] +
  10^{-r}A_{n + 2}
\end{eqnarray*}
it is clear that each new digit found in the $\sqrt[r]{M}$ requires us to work with $r$ more digits (since we need to multiply by $10^r$). This makes the meta-algorithm impractical for manual calculations for large values of $r$. But for cubic roots, the optimized algorithm is still quite practical for manual calculation.

\end{document}